\documentclass[aps,prl,twocolumn,superscriptaddress,showpacs,preprintnumbers,amsmath,amssymb]{revtex4}
\usepackage{graphicx}
\usepackage{dcolumn}
\usepackage{bm}
\usepackage{pstricks}
\usepackage{pst-node}

\newcommand{\pipi}{\pi^+\pi^-}
\newcommand{\pizpiz}{\pi^0\pi^0}

\newcommand{\apipi}{{\cal A}_{\pi\pi}}
\newcommand{\spipi}{{\cal S}_{\pi\pi}}
\newcommand{\apizpiz}{{\cal A}_{\pi^0\pi^0}}
\newcommand{\spizpiz}{{\cal S}_{\pi^0\pi^0}}
\newcommand{\apippim}{{\cal A}_{\pi^+\pi^-}}
\newcommand{\spippim}{{\cal S}_{\pi^+\pi^-}}

\newcommand{\de}{\Delta E}
\newcommand{\mbc}{M_{\rm bc}}
\newcommand{\dt}{\Delta t}
\newcommand{\dmd}{\Delta m_d}

\newcommand{\skstrg}{{\cal S}_{K^*\gamma}}

\usepackage{epsfig}

\usepackage{relsize}
\def\babar{\mbox{\slshape B\kern-0.1em{\smaller A}\kern-0.1em
    B\kern-0.1em{\smaller A\kern-0.2em R}}}

\begin{document}

\preprint{\vbox{ \hbox{   }
                 \hbox{}
}}



\title{\quad\\[0.5cm] \boldmath 
{
New Measurements Using External Photon Conversion at a High Luminosity 
$B$ Factory
}
}

\date{\today}

\affiliation{Department of Physics, Tokyo Institute of Technology, Tokyo}
\affiliation{High Energy Accelerator Research Organization (KEK), Tsukuba}
\affiliation{Department of Physics, Nagoya University, Nagoya}
\author{H.~Ishino}\affiliation{Department of Physics, Tokyo Institute of Technology, Tokyo}
\author{M.~Hazumi} \affiliation{High Energy Accelerator Research Organization (KEK), Tsukuba}
\author{M.~Nakao} \affiliation{High Energy Accelerator Research Organization (KEK), Tsukuba}
\author{T.~Yoshikawa}\affiliation{Department of Physics, Nagoya University, Nagoya}

\begin{abstract}
We propose two novel methods for testing the standard model
using external photon conversion 
at a high-luminosity $e^+e^-$ $B$ factory proposed recently.
The first method is to measure the mixing-induced $CP$-violation
parameter $\spizpiz$ in $B^0\to\pizpiz$ decays.
The precision of $\spizpiz$ is estimated to be 0.23
from a Monte Carlo study for a data sample containing $50 \times 10^9$
$B\overline{B}$ pairs.
We demonstrate that this measurement is crucial 
for reducing the discrete ambiguity of the
Cabibbo-Kobayashi-Maskawa angle $\phi_2$
determined from the isospin analysis with $B\to\pi\pi$ decays.
The second method is to measure 
photon polarization in $B^0\to K^{*0}(\to K^+\pi^-) \gamma$ decays 
using the external photon conversion,
and combine it with $\skstrg$ from $B^0\to K^{*0}(\to K^0_S\pi^0)\gamma$ decays.
This offers a promising way of determining the 
hypothetical right-handed current amplitude and phase 
beyond the standard model.
\end{abstract}


\maketitle

Recent experimental efforts made by two $B$ factory experiments,
BaBar and Belle, have been providing crucial tests of the unitarity of
the Cabibbo-Kobayashi-Maskawa (CKM) matrix~\cite{KM}
in the standard model (SM).
Although no compelling evidence for new physics (NP) beyond the SM
has been found so far,
further accurate tests of the SM quark flavor sector are necessary 
in the next decade in which the NP search will also be extensively performed
at the LHC experiments.

In this Letter, we present two proposals of the novel SM tests
using external photon conversion (PC) that can be carried out only at 
a high luminosity $B$ factory.
Photons from $B$ decays are converted  into $e^+e^-$ pairs by the interaction with the detector at
a certain probability, 
which is known as the Bethe-Heitler process~\cite{bethe-heitler}.
The converted photon provides additional information such as 
the $B$ vertex position and photon polarization.
The photon energy resolution is also improved.
The first proposal is to measure the mixing-induced $CP$-violation
parameter $\spizpiz$ in $B^0\to\pizpiz$ decays~\cite{CC} with PC,
which is otherwise difficult
since the $B^0$ decay vertex cannot be determined. 
We demonstrate that $\spizpiz$ is crucial 
for an unambiguous determination of the CKM 
angle $\phi_2$~\cite{alpha}.
The second proposal is to measure the photon polarization 
in the $B^0\to K^{*0}(\to K^+\pi^-)\gamma$ decays.
This allows us to determine the hypothetical right-handed current
amplitude and phase beyond the SM,
when it is combined with a measurement of the mixing-induced $CP$-violation.
For both proposals, we estimate the precision of the measurements for
a data sample
containing $50\times 10^{9}$ $B\overline{B}$ pairs
(50~ab$^{-1}$ data), which is expected
at a future high luminosity $B$ factory~\cite{superb-loi}.

The angle $\phi_2$
has been measured using $B$ meson decays
into $\pi\pi$~\cite{phi2-pipi-belle, phi2-pipi-babar}.
In the decay $B^0\to\pi\pi$,
where $\pi\pi$ denotes either $\pipi$ or $\pizpiz$, 
$\phi_2$ is obtained from
$\spipi = \sqrt{1-\apipi^2}\sin(2\phi_2 + \kappa_{\pi\pi})$.
Here $\apipi$ is the direct $CP$-violation parameter,
and the phase $\kappa_{\pi\pi}$ can be measured
using isospin relations~\cite{isospin} with the 
branching fractions for $B^0\to \pipi$, $\pizpiz$ and
$B^+\to\pi^+\pi^0$ decays, and 
the direct $CP$-violation parameters
${\cal A}_{\pipi}$ and ${\cal A}_{\pizpiz}$.

In general, we have eightfold ambiguity in the $\phi_2$ solutions
obtained from the isospin analysis in $B\to\pi\pi$ decays
without $\spizpiz$ information.
Measuring ${\cal S}_{\pizpiz}$ can reduce the ambiguity to two,
providing us not only more stringent 
$\phi_2$ constraints but also a severe consistency check
of the $\phi_2$ measurements with $B\to\rho\pi$~\cite{phi2-rhopi}
or $B\to\rho\rho$~\cite{phi2-rhorho}.
The $\pi\pi$ system is free from systematic and theoretical
uncertainties due to the finite width of $\rho$ meson.
The $\spizpiz$ measurement can also probe the $\Delta I=5/2$
contribution in $B\to\pi\pi$ decays~\cite{DeltaI52}.

To estimate the measurement precision of $\spizpiz$,
we employ a Geant detector Monte Carlo (MC) simulation 
developed by the Belle collaboration.
The MC simulation involves the Belle detector~\cite{belle-detector} 
at the KEKB $e^+e^-$
asymmetric-energy collider~\cite{kekb} 
operating at the $\Upsilon(4S)$ resonance produced
with a Lorentz boost factor
of $\beta\gamma=0.425$ along the electron beam direction ($z$ axis).
The Belle detector consists of a 1.5~cm radius beryllium beampipe,
a four-layer silicon vertex detector (SVD) and devices
for tracking, particle identification and 
electromagnetic shower detection.

In the Geant MC simulation, we generate a large number of 
$\Upsilon(4S)\to B^0 \overline{B}{}^0$
decays, where one of the $B^0$ mesons decays into $\pizpiz$ 
and the other decays into a flavor specific state $f_{\rm tag}$.
The time dependent decay rate is~\cite{sanda}
\begin{eqnarray}
{\cal P}^q_{\pizpiz}(\dt) & = & 
      \frac{e^{-|\dt|/\tau_{B^0}}}{4\tau_{B^0}}\{1+q(1-2w)   \label{eq:tcpv} \\
                & & [\spizpiz\sin(\dmd\dt) 
                    + \apizpiz\cos(\dmd\dt)]\}, \nonumber
\end{eqnarray}
where $\tau_{B^0}$ is the $B^0$ lifetime, $\dmd$ is the mass difference
between the two $B^0$ mass eigenstates, $\dt = t_{CP} - t_{\rm tag}$,
$t_{CP}$ ($t_{\rm tag}$) is the decay time of $B^0\to\pizpiz$ ($f_{\rm tag}$),
$q=+1$ ($-1$) when $f_{\rm tag}=B^0$ ($\overline{B}{}^0$),
and $w$ is the wrong tag fraction.
Since the two $B$ mesons are produced nearly at rest in the 
$\Upsilon(4S)$ center-of-mass system (CMS),
$\dt$ is determined from the distance
between the two $B$ meson decay vertices along the 
$z$-direction ($\Delta z$);
$\Delta t \cong \Delta z / c\beta\gamma$, where $c$ is the speed of light.
In the MC, we find 11.3\% of events have photons
converted to $e^+e^-$ pairs at the beampipe or the SVD, 
while 2.2\% of events have the $\pi^0$ Dalitz decay.

By combining $\pi^0\to\gamma\gamma$ and $\pi^0\to\gamma e^+e^-$,
we choose the $B^0\to\pizpiz$ signal candidates
using the energy difference $\Delta E = E^*_B - E^*_{\rm beam}$
and the beam-energy constrained mass 
$M_{\rm bc} = \sqrt{(E^*_{\rm beam})^2 - (p^*_B)^2}$,
where $E^*_{\rm beam}$ is the CMS beam-energy,
and $E^*_B$ and $p^*_B$ are the CMS energy and momentum of the candidate.
For the vertex reconstruction, 
we require one of the two tracks of the $e^+e^-$ pair to have
at least two associated hits in the SVD.

The decay vertex reconstruction technique in Ref.~\cite{ksvtx}
utilizing $K^0_S\to\pipi$ and the $e^+e^-$ beam interaction profile
is employed for the candidates with the PC,
while the algorithm~\cite{normvtx} is applied to 
the $B^0\to\pizpiz$ candidates 
followed by the $\pi^0$ Dalitz decay and the $f_{\rm tag}$ state.
The reconstruction efficiency of 
$B^0\to\pizpiz$ involving the PC ($\pi^0$ Dalitz decay)
is 8.6\% (14.8\%), leading to the expected signal yield of 690 (230)
for the branching fraction 
in Table~\ref{table:pipi-input}~\cite{hfag2006, note-for-table}.

\begin{table}
\begin{center}
\caption{
Branching fractions and $CP$-violation parameters
for $B\to\pi\pi$ decays used in our study.
The first (second) errors represent statistical (systematic)
uncertainties.
}
\begin{tabular}{cc}
\hline\hline
${\cal B}(B^0\to\pipi)$      & $(5.21\pm0.02\pm0.10)\times 10^{-6}$ \\
${\cal B}(B^+\to\pi^+\pi^0)$ & $(5.61\pm0.04\pm0.17)\times 10^{-6}$ \\
${\cal B}(B^0\to \pizpiz)$   & $(1.35\pm0.02\pm0.05)\times 10^{-6}$ \\
$\spippim$                   & $-0.66\pm0.01\pm0.01$ \\
$\apippim$                   & $+0.37\pm0.01\pm0.01$ \\
$\spizpiz$                   & $+0.92$ \\
$\apizpiz$                   & $+0.16\pm0.03\pm0.01$ \\
\hline\hline
\end{tabular}
\label{table:pipi-input}
\end{center}
\end{table}

Based on the residuals between MC-generated and reconstructed positions,
we find the resolutions of the $B^0\to\pizpiz$ decay position
and $\Delta z$ are approximately $120$~$\mu$m and $150$~$\mu$m, respectively.
From the $\Delta z$ residual distribution, a resolution function
of $\dt$ is constructed.
We model it as a sum of three Gaussians.
To obtain a $\dt$ probability density function (PDF), 
we convolve the resolution function with the time-dependent decay rate
in Eq.~(\ref{eq:tcpv}).
We assume the effective tagging efficiency $\varepsilon= 30$\%, i.e. 
$w=(1-\sqrt{\varepsilon})/2=0.23$~\cite{flavor-tag}.

We also take into account 
the continuum $e^+e^-\to q\overline{q}$
$(q = u, d, s, c)$ and $B^+\to\rho^+\pi^0$ rare decays~\cite{B2pi0pi0}.
Using a large Geant MC sample, 
the expected yield of the continuum ($B^+\to\rho^+\pi^0$) background 
is estimated to be 20000 (300). 
We employ the $\dt$ PDF for the backgrounds in Ref.~\cite{phi2-pipi-belle},
which contains prompt and lifetime components convolved with
a resolution function composed of a sum of two Gaussians.

To distinguish between the signal and the continuum events,
we construct a likelihood function ${\cal L}_{S}$ (${\cal L}_{BG}$)
for the signal (continuum) events
from the event topology and the $B$ flight direction in the CMS
with respect to the $z$ axis, and form a likelihood ratio
${\cal R}={\cal L}_{S}/({\cal L}_{S}+{\cal L}_{BG})$
for the candidate events.

We generate 2000 MC pseudo experiments with the input values of
$\spizpiz$ and $\apizpiz$ listed in Table~\ref{table:pipi-input};
each pseudo experiment contains the signal, 
continuum and $B^+\to\rho^+\pi^0$ events
with the yields estimated above.
The MC events are generated according to the 
PDFs of $\de$, $\mbc$ and ${\cal R}$, which are
determined from the Geant MC.
The $\dt$ PDFs defined above are 
also used for the MC generation.

For a fit to obtain $\spizpiz$, 
a likelihood value of the $i$-th event is defined as
$
P_i = \sum_k n_k 
{\cal P}_k(\vec{s}_i) {\cal P}_k(\dt_i),
$
where $n_k$ is the fraction of component $k$
indicating either signal, continuum or $B^+\to\rho^+\pi^0$,
${\cal P}_k(\vec{s})$ is the event-by-event probability
for the component $k$
as a function of $\vec{s} = (\de, \mbc, {\cal R})$,
and ${\cal P}_k(\dt)$ is the $\dt$ PDF of component $k$. 
We obtain $\spizpiz$ and $\apizpiz$ simultaneously
by maximizing the likelihood function ${\cal P} = \prod_i P_i$
in each pseudo experiment.
The expected $\spizpiz$ error $\sigma_{\spizpiz}$ is determined
from a root mean square value of the $\spizpiz$ distribution;
we measure $\sigma_{\spizpiz}=0.23$.

\begin{figure}
\epsfxsize=8cm  \epsfbox{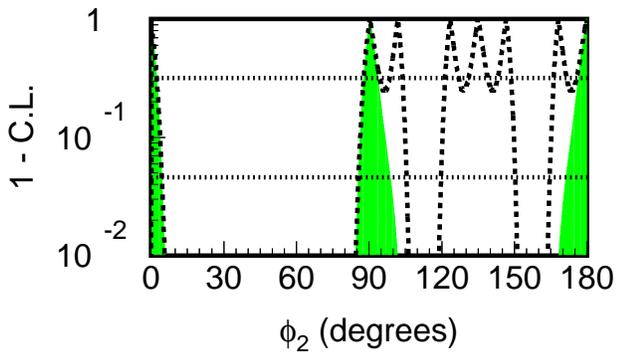}
\caption{
Confidence level as a function of $\phi_2$ obtained from the
isospin analysis.
The filled (dotted) curve shows the C.L. with (without) 
the $\spizpiz$ constraint.
Two dashed horizontal lines indicate C.L. = 68.3\% and 95.4\%.
}
\label{fig:phi2-constraint}
\end{figure}

To constrain $\phi_2$, we perform an isospin analysis
using the obtained $\sigma_{\spizpiz}$ value
and the values in Table~\ref{table:pipi-input} 
with the statistical approach described in~\cite{r-fit}.
The $\spizpiz$ central value is obtained from the isospin relations
by assuming $\phi_2=90^{\circ}$.
The statistical errors in Table~\ref{table:pipi-input}
are estimated by multiplying 0.1 to the errors in Ref.~\cite{hfag2006}
assuming that statistics are 100 times 
as large as those in Ref.~\cite{hfag2006}.
The systematic errors of the
branching fractions are assumed to 
arise from uncertainties in the detection efficiency for
a $\pi^0$ (2\%) and a charged particle (1\%)~\cite{hh-belle}.
We assume 1\% systematic errors for the $CP$-violation parameters,
which originate from the asymmetry of charged 
particle detection efficiency~\cite{hh-belle}
and the irreducible vertex reconstruction uncertainty of
SVD mis-alignment~\cite{superb-loi}.
Figure~\ref{fig:phi2-constraint} shows the obtained
confidence levels (C.L.) as a function of $\phi_2$ 
with and without the $\spizpiz$ constraint.
While eight-fold discrete ambiguity is seen in the case without $\spizpiz$,
it reduces to two by including the $\spizpiz$ constraint.

We now turn to the photon polarization measurement with PC.
The SM predicts that 
the helicity of an emitted photon 
in the radiative transition $b\to s\gamma$
is dominantly left-handed. 
Therefore the detection of a right handed photon is unambiguous
evidence for NP.
The external PC in $B^0\to K^{*0}\gamma$ followed
by $K^{*0}\to K^+\pi^-$ enables us to measure the amplitude of
the right-handed photon emission by making use of
the photon polarization information obtained
from the angle $\phi$ between the event planes of 
$K^+\pi^-$ and $\gamma\to e^+e^-$~\cite{grossman-pirjol}.

The amplitude for the emission of a left- (right-) handed photon
is expressed as $F_L = M e^{i\phi_L}\cos\psi$ 
$(F_R = M e^{i\phi_R}\sin\psi)$,
where $\phi_L$ ($\phi_R$) is a $CP$-violating phase,
$\psi$ is ${\cal O}(m_s/m_b)$ in the SM~\cite{footnote1}, $m_s$ $(m_b)$ is
the $s$ $(b)$ quark mass, and $M$ is an amplitude
that determines the overall decay rate.
The distribution for the angle $\phi$ satisfies
\begin{eqnarray}
\frac{d\sigma}{d\phi} \propto 1 + \xi R \cos(2\phi + \delta), 
\label{eq:phi-distribution} \\
R = \frac{|F_L||F_R|}{|F_L|^2+|F_R|^2} = \frac{1}{2}\sin2\psi,
\end{eqnarray}
where $\xi = \sqrt{X^2+Y^2}$,
$\tan\delta = (X\tan\phi_- - Y)/(X+Y\tan\phi_-)$,
$\phi_- = \phi_R - \phi_L$,
$X = 2(\sigma_{\rm II} - \sigma_{\rm III}) /
(\sigma_{\rm II}+\sigma_{\rm III})$ and
$Y = 4\sigma_{\rm IV}/(\sigma_{\rm II}+\sigma_{\rm III})$.
Here $\sigma_{\rm II}$ ($\sigma_{\rm III}$) is the PC cross section
parallel (perpendicular) to the polarization 
of a polarized photon~\cite{gluckstern}, and
$\sigma_{\rm IV}$ measures the acoplanarity of the photon, electron and
positron vectors.
We ignore the contribution from $\sigma_{\rm IV}$ 
by following Ref.~\cite{grossman-pirjol}, hence
assume $\delta = \phi_-$.
The dilution parameter $\xi$ depends on both photon energy
and an opening angle of the $e^+e^-$ pair, and is approximately 0.1
when integrated over the electron energy 
and the opening angle~\cite{pc-dilution}.
From the $\phi$ distribution, we can measure $\delta$ and $R$ 
with $B^0\to K^{*0}(\to K^+\pi^-)\gamma$ decays.

In addition, the time-dependent $CP$ asymmetry measurements
in $B^0\to K^{*0}\gamma$ followed by $K^{*0}\to K_S^0\pi^0$~\cite{bsg}
yield the mixing-induced $CP$-violation parameter
\begin{equation}
\skstrg = -2R\sin(2\phi_1 - \phi_+),
\label{eq:skstrgamma}
\end{equation}
where $\phi_1=(21.7\pm1.3)^{\circ}$~\cite{hfag2006} is 
one of the CKM weak phases,
and $\phi_+ = \phi_R + \phi_L$.
We point out that
we can determine $R$, $\phi_L$ and $\phi_R$ separately
by combining the polarization and $CP$ asymmetry measurements.
Within the SM, we expect $\phi_L = \phi_R = 0$ to a good
approximation~\cite{Atwood:2007qh}. 
Observation of the $CP$-violating phases 
would thus be a clear evidence for NP.

We generate a large number of 
$\Upsilon(4S) \to B^0\overline{B}{}^0 \to (K^{*0}\gamma)(f_{\rm tag})$
events, where $K^{*0}$ decays into $K^+\pi^-$ 
in the Geant MC simulation.
The photon emitted is converted at the beampipe
or the SVD with a probability of 2.8\%.
The $B^0$ reconstruction efficiency is 0.36\% including the
branching fraction of $K^{*0}\to K^+\pi^-$; 
we expect 7200 $B^0\to K^{*0}\gamma$ events having
converted photons
with the branching fraction of $4\times 10^{-5}$~\cite{hfag2006}.
Because of the small opening angle of the $e^+e^-$ pair, typically 10~mrad,
we find the $\phi$ resolution is $23^{\circ}$ 
and the $\phi$ reconstruction efficiency is 35\%; 
the remaining 65\% of events have no information on $\phi$
and produce a flat $\phi$ distribution.

We estimate the expected measurement precision of 
$R$ and $\phi_-$.
In this study we ignore the possible background contributions of
about 5\%, 
which is estimated using a large Geant MC sample.
We generate 1000 pseudo experiments for
five $R$ values from 0.1 to 0.5, 
while $\phi_-$ is fixed to 0.
Each pseudo experiment contains 7200 signal events
generated according to Eq.~(\ref{eq:phi-distribution})
modified to take into account the $\phi$ reconstruction efficiency and
resolution.
We choose $x=R\cos(\phi_-)$ and $y=R\sin(\phi_-)$ 
as fit parameters,
and find the $x$ and $y$ distributions
have the same Gaussian sigma $\sigma_x = 0.52$.
The result is independent of $R$ values,
and neither bias nor correlation is found.
The same procedure is performed with 72000 signal events per
pseudo experiment, corresponding to a data sample 
containing $500\times 10^9$ $B\overline{B}$ pairs (500~ab$^{-1}$ data).
We obtain $\sigma_x=0.16$,
consistent with the expectation $0.52/\sqrt{10}$.

\begin{figure}
\epsfxsize=6.5cm  \epsfbox{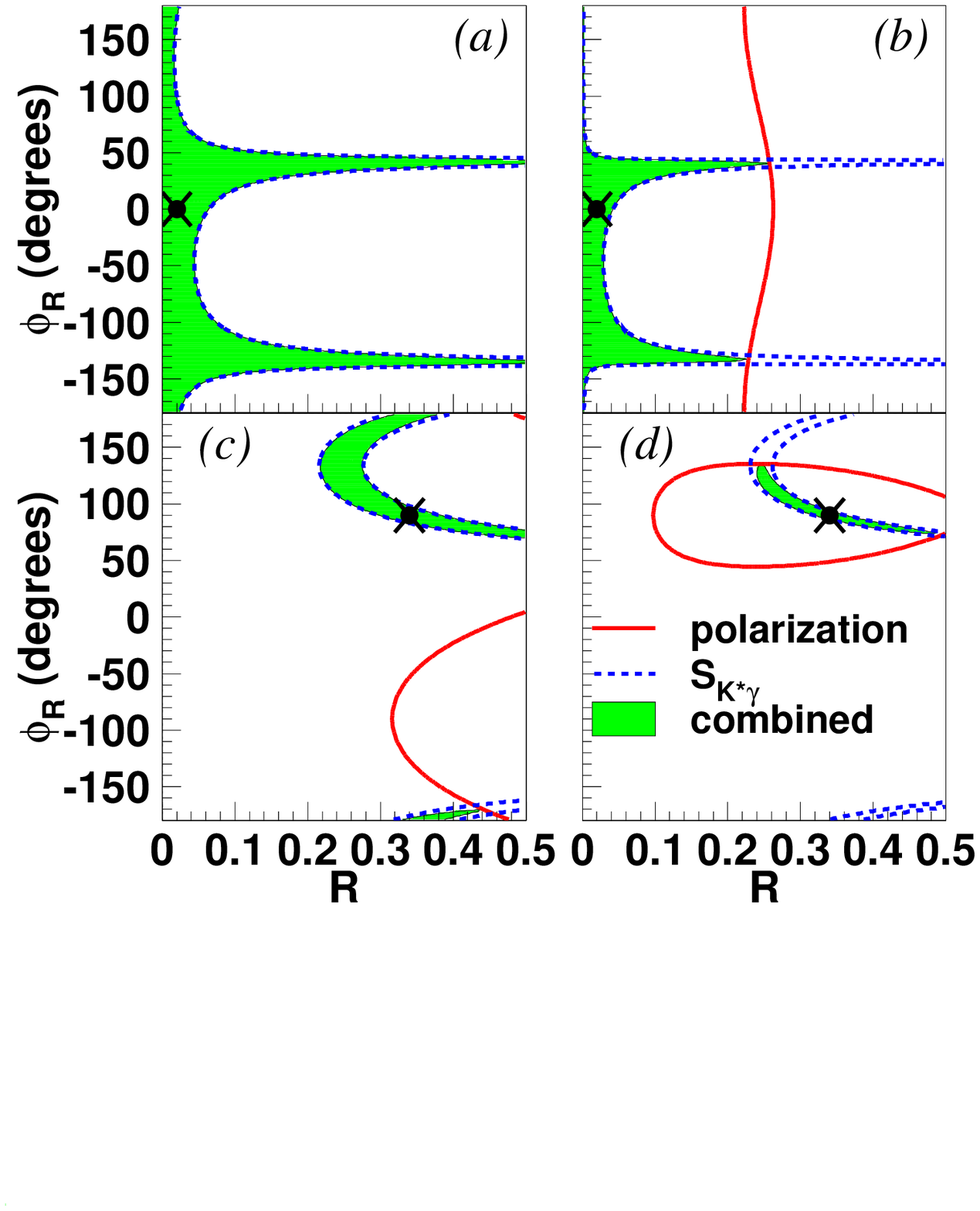}
\caption{
Confidence regions in $\phi_R$ {\sl vs.} $R$ plane.
The contours indicate C.L.$=68.3$\% ($1\sigma$).
The cross marks indicate the assumed ($R$, $\phi_R$) positions;
figures (a) and (b) correspond to (0.02, $0^{\circ}$), 
while figures (c) and (d) correspond to (0.34, $90^{\circ}$).
Left (right) two figures show the sensitivity with the
50 (500)~ab$^{-1}$ data.
}
\label{fig:kpigamma}
\end{figure}

To further constrain $R$ and the phases,
we make use of Eq.~(\ref{eq:skstrgamma}).
Since $F_L$ is dominated by the SM contribution,
we fix $\phi_L$ to 0;
hence $\phi_- = \phi_+ = \phi_R$.
We have two independent constraints on $x$ and $y$
from the polarization $\phi$ measurement and 
$\skstrg = -2x\sin2\phi_1+2y\cos2\phi_1$.
The measurement precision of $\skstrg$ is expected to be
$0.04$ ($0.02$) for the 50 (500)~ab$^{-1}$ data~\cite{superb-loi}.
For the sensitivity estimation,
we employ a frequentist statistical approach in Ref.~\cite{fc}.
We examine two cases: the SM expectation ($R$, $\phi_R$) = (0.02, $0^{\circ}$)
and the left-right symmetric model assumption
(0.34, $90^{\circ}$)~\cite{bsg,lrsm}.
Figure~\ref{fig:kpigamma} shows the obtained confidence regions
in $\phi_R$ {\sl vs.} $R$ plane.
With the 50~ab$^{-1}$ data, the constraint is mostly determined by
the $\skstrg$ measurement precision because of the large $\sigma_x$ value.
On the other hand, with the 500~ab$^{-1}$ data the $\phi$ measurement becomes
important for the $R$ and $\phi_R$ constraints. 

We emphasize 
the vertex reconstruction using PC
can also be applied to the time-dependent $CP$ violation
analysis in $B^0\to K^0\pi^0$ decays, which are sensitive
to a new $CP$-violating phase in the $b\to s\bar{q}q$ transition; 
we can increase
the vertex efficiency by about 10\% for $B^0\to K^0_S\pi^0$
and can reconstruct the vertex position of $B^0\to K^0_L\pi^0$.
The energy resolution of a converted photon
is three times better than that of a photon without conversion.
This feature improves the signal-to-noise ratio
in the exclusive $b\to s\gamma$ measurements,
and in the search for the lepton number violating process
$\tau\to\mu\gamma$.

Finally, we note that the PC probability increases by 50\%
if we use a six-layer SVD proposed in~\cite{superb-loi}.
Even a higher PC probability is possible by choosing high-$Z$ material,
such as CdTe, for the SVD. Dedicated studies are needed in this case
to guarantee good momentum resolution for charged particles.

In summary, we have made two proposals of new measurements
using PC at a future high luminosity $B$ factory experiment.
The PC enables us to determine the vertex
position of $B^0\to\pizpiz$ decays. 
With the 50~ab$^{-1}$ data, 
we find that the measurement precision of $\spizpiz$ is 0.23, and that
it reduces the discrete ambiguity of the $\phi_2$ solutions.
The photon polarization measured using PC
combined with the $\skstrg$ measurement in $B^0\to K^{*0}\gamma$ decays
allows us to constrain the phase and amplitude of the right-handed
current beyond the SM.
We find that with the 500~ab$^{-1}$ data sample
the polarization measurement
becomes important for constraining $R$ and $\phi_R$.

We thank the Belle collaboration for providing the detector Geant MC 
simulation.
M.H. is supported in part by 
a JSPS Grant-in-Aid for Scientific Research (C), 17540279.


\clearpage
\newpage


\begin{thebibliography}{999}
\bibitem{CC}
Throughout this paper, 
the inclusion of the charge conjugate mode decay is implied
unless otherwise stated.
\bibitem{alpha} Another naming convention, 
$\alpha$ ($= \phi_2$), 
is also used in the literature.
\end{thebibliography}
\end{document}